\documentclass{sig-alternate-05-2015}

\usepackage{xspace}
\usepackage{hyperref}
\usepackage{cleveref} %
\usepackage{tikz}
\usepackage{flushend} %
\usepackage{caption,setspace}
\usepackage[binary-units]{siunitx}
\DeclareSIUnit{\packet}{p}

\newcommand{\etal}{et al.\xspace}
\newcommand{\ie}{i.e.,\xspace}
\newcommand{\eg}{e.g.,\xspace}
\newcommand{\ixp}{large European IXP\xspace}
\newcommand{\todo}[1]{%
\ifstatus
        \textcolor{red}{\textbf{TODO:} #1\xspace}
\fi
}

\newcommand{\para}[1]{\noindent\textbf{#1}\xspace}

\DeclareSIUnit{\nothing}{\relax}

\newcommand{\mn}{\mega\nothing}
\newcommand{\kn}{\kilo\nothing}

\newcommand{\sk}[1]{\SI{#1}{\kn}}
\newcommand{\sm}[1]{\SI{#1}{\mn}}

\newcommand{\sperc}[1]{\SI{#1}{\percent}}

\def\sharedaffiliation{%
\end{tabular}
\begin{tabular}{c}%
}

\begin{document}

\newif\ifstatus
\statustrue
\statusfalse

\conferenceinfo{Passive and Active Measurement Conference 2020}{March 30--31, 2020, Eugene, OR, USA}

\title{Reserved: Dissecting Internet Traffic on Port 0}

\numberofauthors{3} %

\author{%
Aniss Maghsoudlou \and Oliver Gasser \and Anja Feldmann
\and
\sharedaffiliation
       \affaddr{Max Planck Institute for Informatics}\\
       \email{\{aniss,oliver.gasser,anja\}@mpi-inf.mpg.de}
}

\maketitle

\begin{abstract}
    Transport protocols use port numbers to allow connection multiplexing on Internet hosts.
    TCP as well as UDP, the two most widely used transport protocols, have limitations on what constitutes a valid and invalid port number.
    One example of an invalid port number for these protocols is port 0.

    In this work, we present preliminary results from analyzing port 0 traffic at a \ixp.
    In one week of traffic we find 74GB port 0 traffic.
    The vast majority of this traffic has both source and destination ports set to 0, suggesting scanning or reconnaissance as its root cause.
    Our analysis also shows that more than half of all port 0 traffic is targeted to just 18 ASes, whereas more than half of all traffic is originated by about 100 ASes, suggesting a more diverse set of source ASes.
\end{abstract}

\section{Introduction}

Port numbers allow a network host to serve multiple applications or services under the same IP address.
When using TCP or UDP as a transport protocol, port numbers are encoded as 16 bit values.
Depending on the used transport protocol, certain ranges of port numbers are assigned to well-known applications (\eg TCP port 443 is assigned to the well-known HTTPS protocol). Another part of port ranges are private ports which are not assigned to well-known protocols, and reserved ports which should not be used at all\cite{ianaports}.
In 1983, Reynolds and Postel declared port 0 as being reserved in RFC~870 \cite{rfc870}; thus disallowing the use of source or destination port 0 in any TCP segment or UDP datagram.

Contrary to this requirement, packets with port 0 are still sent through the public Internet as shown by Bou-Harb \etal \cite{bou2014multidimensional}.
In this research, we expand upon previous work by investigating port 0 traffic at a \ixp.
In addition, we combine passive measurements with data obtained from active measurements to identify IXP traffic destined or originating from web servers \cite{rapid7}.

\section{Datasets}

In our research, we leverage two data sources: passive measurement data obtained at a \ixp and active measurement results from Rapid7 \cite{rapid7}.

At the IXP we capture IPFIX flow data with packet sampling applied resulting in 1 out of every \sk{10} packets being sampled.
Our results are based on one week of IPv4 flow data between September 1, 2019 and September 7, 2019.
Due to its nature, sampled flow data does not provide a complete view of network traffic by under-representing the number of flows \cite{brauckhoff2006impact}.
Nevertheless, it allows us to analyze traffic characteristics of port 0 traffic such as the number of bytes and packets and providing a lower bound on the number of flows.
For our analysis, we load the obtained flow data into the ClickHouse DBMS \cite{clickhouse}.

In addition to the passive dataset from the IXP, we also use results from active measurements.
Specifically, we leverage Rapid7's Project Sonar \cite{rapid7} to identify IP addresses found in port 0 traffic as HTTP and HTTPS servers.

\section{Preliminary Findings}

We observed that out of the complete \SI{29}{\tera\byte} of traffic consisting of 42 billion packets, \SI{74}{\giga\byte} (in 100 million packets, \ie \sperc{0.24} of all seen packets) have either the source or destination port set to 0.
More than \sperc{99} of the sampled port 0 packets have both source and destination ports set to 0.

\begin{figure}[htb]
  \captionsetup{skip=.25em}
        \centering
        \includegraphics[width=\columnwidth]{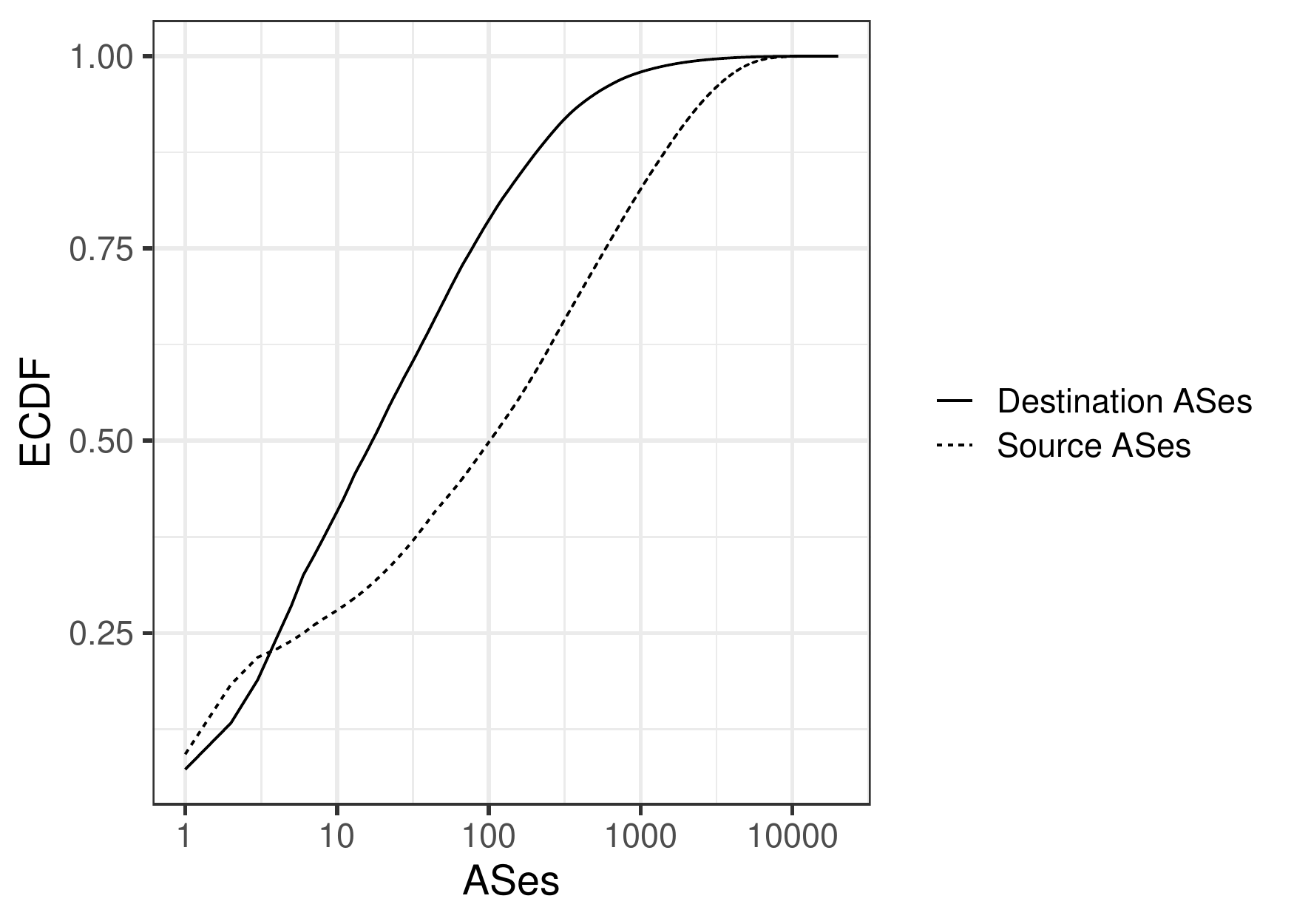}
        \caption{Cumulative Distribution of ASes involved in port 0 traffic. Note that the x axis is log-scaled.}
        \label{fig:cdf}
\end{figure}

When mapping the involved source and destination IP address to ASes \cite{rv}, we find that the AS distribution is very top heavy.
We observed port 0 traffic in \sk{180} prefixes from \num{24654} ASes.
\Cref{fig:cdf} shows that the vast majority of port 0 traffic originates from and is destined to a small number of ASes.

\begin{figure}[htb]
  \captionsetup{skip=.25em}
        \centering
        \includegraphics[width=\columnwidth]{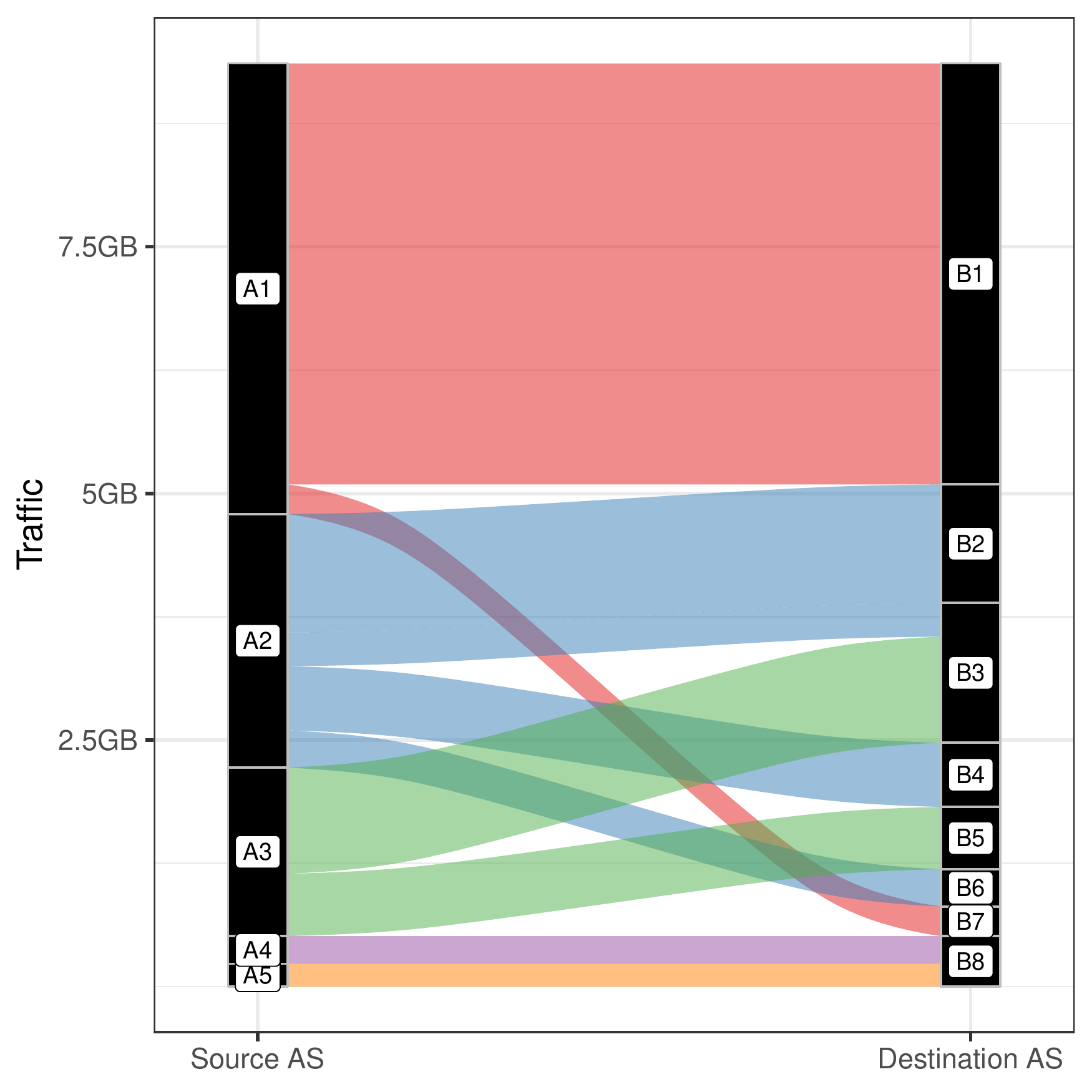}
        \caption{Traffic between top 10 (source AS, destination AS) pairs involved in port 0 traffic.}
        \label{fig:alluvial}
\end{figure}

Next, we delve into the port 0 flows to better understand source and destination ASes.
\Cref{fig:alluvial} shows port 0 traffic from source to destination AS.
We see that a large portion of port 0 flows between only two ASes, originating from a cloud VM provider and destined to a Central American ISP.

\todo{How much of overall traffic from these ASes does port 0 traffic represent? Much more compared to other ASes?}

\todo{Top X ASes with largest port 0 traffic percentage}

To better understand the hosts behind port 0 traffic, we combine our passive data with active measurement results.
We use Rapid7's TCP/80 and TCP/443 measurements to classify HTTP and HTTPS servers.
By combining the active data and the IPFIX data, we observe that out of \sm{2.6} total IP addresses involved in port 0 traffic, about \sk{308.3} (\sperc{12}) belong to web servers.
These server IP addresses were involved in port 0 traffic as both source IP and destination IP.

Finally, we also evaluate whether port 0 traffic is bidirectional.
Out of \sm{17.2} source and destination IP address pairs, we only observe \sk{25.0} (\sperc{0.14}) in the reverse order, which we interpret as bidirectional.
When evaluating bidirectional traffic for web servers, we find that of the \sk{14.8} IP addresses involved in bidirectional port 0 traffic, only \sk{3.4} are identified as web servers.

\section{Conclusion and Future Work}

In this work we used passive data from a \ixp to analyze the use of port 0 in Internet traffic.
Even though the overall share of port 0 traffic was quite small, we found that a small number of ASes contributes a large share of port 0 traffic.
When combining the passive measurements with results from active measurements, we saw that more than \sperc{10} of port 0 IP addresses also run a web server.
These servers could be either targets or sources of port 0 vulnerability scanning.

In the future, we plan to perform active measurements to better understand how many networks filter port 0 compared to regular traffic.
Furthermore, we strive to analyze longer timespans of IXP flow data.
Finally, we also plan to investigate at port 0 traffic in IPv6 traffic and use active measurement results for identifying port 0 traffic on IPv6 servers \cite{gasser2018clusters}.

\para{Acknowledgments:} We thank the \ixp for providing the flow data and Rapid7 for their publicly available measurement results.

\bibliographystyle{abbrv}
\bibliography{references}

\end{document}